\documentclass[aps,prl,reprint,superscriptaddress,showpacs]{revtex4-1}

\usepackage{graphicx}  
\usepackage{subfigure}

\begin{document}

\title{Measurement of the Atmospheric $\nu_e$ flux in IceCube}

\affiliation{III. Physikalisches Institut, RWTH Aachen University, D-52056 Aachen, Germany}
\affiliation{School of Chemistry \& Physics, University of Adelaide, Adelaide SA, 5005 Australia}
\affiliation{Dept.~of Physics and Astronomy, University of Alaska Anchorage, 3211 Providence Dr., Anchorage, AK 99508, USA}
\affiliation{CTSPS, Clark-Atlanta University, Atlanta, GA 30314, USA}
\affiliation{School of Physics and Center for Relativistic Astrophysics, Georgia Institute of Technology, Atlanta, GA 30332, USA}
\affiliation{Dept.~of Physics, Southern University, Baton Rouge, LA 70813, USA}
\affiliation{Dept.~of Physics, University of California, Berkeley, CA 94720, USA}
\affiliation{Lawrence Berkeley National Laboratory, Berkeley, CA 94720, USA}
\affiliation{Institut f\"ur Physik, Humboldt-Universit\"at zu Berlin, D-12489 Berlin, Germany}
\affiliation{Fakult\"at f\"ur Physik \& Astronomie, Ruhr-Universit\"at Bochum, D-44780 Bochum, Germany}
\affiliation{Physikalisches Institut, Universit\"at Bonn, Nussallee 12, D-53115 Bonn, Germany}
\affiliation{Universit\'e Libre de Bruxelles, Science Faculty CP230, B-1050 Brussels, Belgium}
\affiliation{Vrije Universiteit Brussel, Dienst ELEM, B-1050 Brussels, Belgium}
\affiliation{Dept.~of Physics, Chiba University, Chiba 263-8522, Japan}
\affiliation{Dept.~of Physics and Astronomy, University of Canterbury, Private Bag 4800, Christchurch, New Zealand}
\affiliation{Dept.~of Physics, University of Maryland, College Park, MD 20742, USA}
\affiliation{Dept.~of Physics and Center for Cosmology and Astro-Particle Physics, Ohio State University, Columbus, OH 43210, USA}
\affiliation{Dept.~of Astronomy, Ohio State University, Columbus, OH 43210, USA}
\affiliation{Dept.~of Physics, TU Dortmund University, D-44221 Dortmund, Germany}
\affiliation{Dept.~of Physics, University of Alberta, Edmonton, Alberta, Canada T6G 2G7}
\affiliation{D\'epartement de physique nucl\'eaire et corpusculaire, Universit\'e de Gen\`eve, CH-1211 Gen\`eve, Switzerland}
\affiliation{Dept.~of Physics and Astronomy, University of Gent, B-9000 Gent, Belgium}
\affiliation{Dept.~of Physics and Astronomy, University of California, Irvine, CA 92697, USA}
\affiliation{Laboratory for High Energy Physics, \'Ecole Polytechnique F\'ed\'erale, CH-1015 Lausanne, Switzerland}
\affiliation{Dept.~of Physics and Astronomy, University of Kansas, Lawrence, KS 66045, USA}
\affiliation{Dept.~of Astronomy, University of Wisconsin, Madison, WI 53706, USA}
\affiliation{Dept.~of Physics and Wisconsin IceCube Particle Astrophysics Center, University of Wisconsin, Madison, WI 53706, USA}
\affiliation{Institute of Physics, University of Mainz, Staudinger Weg 7, D-55099 Mainz, Germany}
\affiliation{Universit\'e de Mons, 7000 Mons, Belgium}
\affiliation{T.U. Munich, D-85748 Garching, Germany}
\affiliation{Bartol Research Institute and Department of Physics and Astronomy, University of Delaware, Newark, DE 19716, USA}
\affiliation{Dept.~of Physics, University of Oxford, 1 Keble Road, Oxford OX1 3NP, UK}
\affiliation{Dept.~of Physics, University of Wisconsin, River Falls, WI 54022, USA}
\affiliation{Oskar Klein Centre and Dept.~of Physics, Stockholm University, SE-10691 Stockholm, Sweden}
\affiliation{Department of Physics and Astronomy, Stony Brook University, Stony Brook, NY 11794-3800, USA}
\affiliation{Dept.~of Physics and Astronomy, University of Alabama, Tuscaloosa, AL 35487, USA}
\affiliation{Dept.~of Astronomy and Astrophysics, Pennsylvania State University, University Park, PA 16802, USA}
\affiliation{Dept.~of Physics, Pennsylvania State University, University Park, PA 16802, USA}
\affiliation{Dept.~of Physics and Astronomy, Uppsala University, Box 516, S-75120 Uppsala, Sweden}
\affiliation{Dept.~of Physics, University of Wuppertal, D-42119 Wuppertal, Germany}
\affiliation{DESY, D-15735 Zeuthen, Germany}

\author{M.~G.~Aartsen}
\affiliation{School of Chemistry \& Physics, University of Adelaide, Adelaide SA, 5005 Australia}
\author{R.~Abbasi}
\affiliation{Dept.~of Physics and Wisconsin IceCube Particle Astrophysics Center, University of Wisconsin, Madison, WI 53706, USA}
\author{Y.~Abdou}
\affiliation{Dept.~of Physics and Astronomy, University of Gent, B-9000 Gent, Belgium}
\author{M.~Ackermann}
\affiliation{DESY, D-15735 Zeuthen, Germany}
\author{J.~Adams}
\affiliation{Dept.~of Physics and Astronomy, University of Canterbury, Private Bag 4800, Christchurch, New Zealand}
\author{J.~A.~Aguilar}
\affiliation{D\'epartement de physique nucl\'eaire et corpusculaire, Universit\'e de Gen\`eve, CH-1211 Gen\`eve, Switzerland}
\author{M.~Ahlers}
\affiliation{Dept.~of Physics and Wisconsin IceCube Particle Astrophysics Center, University of Wisconsin, Madison, WI 53706, USA}
\author{D.~Altmann}
\affiliation{Institut f\"ur Physik, Humboldt-Universit\"at zu Berlin, D-12489 Berlin, Germany}
\author{J.~Auffenberg}
\affiliation{Dept.~of Physics and Wisconsin IceCube Particle Astrophysics Center, University of Wisconsin, Madison, WI 53706, USA}
\author{X.~Bai}
\thanks{Physics Department, South Dakota School of Mines and Technology, Rapid City, SD 57701, USA}
\affiliation{Bartol Research Institute and Department of Physics and Astronomy, University of Delaware, Newark, DE 19716, USA}
% ...

\date{\today}

\collaboration{IceCube Collaboration}
\noaffiliation

\begin{abstract}
  We report the first measurement of the atmospheric electron neutrino flux in the energy range 
  between approximately 80~GeV and 6~TeV, using data recorded during the first year of operation 
  of IceCube's DeepCore low energy extension.
  Techniques to identify neutrinos interacting within the DeepCore volume and veto muons originating outside the detector are demonstrated.
  A sample of 1029 events is observed in 281 days of data, 
  of which $496 \pm 66 \text{(stat.)} \pm \text{88(syst.)}$ are estimated to be cascade events, 
  including both electron neutrino and neutral current events.  
  The rest of the sample includes residual backgrounds due to atmospheric muons and charged current 
  interactions of atmospheric muon neutrinos.  The flux of the atmospheric electron neutrinos is 
  consistent with models of atmospheric neutrinos in this energy range.  
  This constitutes the first observation of electron neutrinos and neutral current interactions 
  in a very large volume neutrino telescope optimized for the TeV energy range.  
\end{abstract}


\maketitle

The atmospheric $\nu_\mu$ spectrum (we do not differentiate between $\overline\nu$ and $\nu$)
has been measured at energies up to 400~TeV~\cite{Diffuse-Warren-2011}.
Much less is known about atmospheric $\nu_e$.
These $\nu_e$ come mostly from the decays of kaons and muons produced in cosmic-ray air showers.
Underground water Cherenkov telescopes like IMB-3 and Super-Kamiokande as well as calorimetric detectors such as
Fr\'ejus, NUSEX, and Soudan-2 have studied atmospheric $\nu_e$ with energies up
to a few tens of~GeV \cite{IMB-Becker-1992,SK-Wendell-2010,Frejus-Daum-1995,NUSEX-Aglietta-1988,Soudan2-Sanchez-2003},
but no measurement has been made at higher energies.
So far, searches for $\nu_e$ at higher energies have yielded upper
limits \cite{Ackermann:2004zw,Cascade-Mike-2011, Cascade-Joanna-2011,Cascade-ICRC-2011}.
Theoretical calculations for atmospheric $\nu_e$ are poorly constrained at energies above 100~GeV
due to the uncertainties in kaon production~\cite{Honda-2007, BartolError-2006}.

In this Letter, we report on a measurement of atmospheric neutrino-induced cascades using the DeepCore infill array in IceCube.
``Cascades'' are $\nu_e$ charged current~(CC) interactions and neutral current~(NC) interactions of neutrinos of all flavors.
From a selected sample of cascade candidate events, an atmospheric $\nu_e$ flux is obtained in the energy range between 80~GeV and 6~TeV.

In addition to their interest for understanding particle production in air showers,
the results presented here are a first step toward measurement of the appearance of $\nu_\tau$
(which is indistinguishable from $\nu_e$ in IceCube at these energies) due to neutrino flavor oscillations~\cite{DC-Koskinen-2011}.  
The first oscillation maximum for $\nu_\mu \rightarrow \nu_\tau$, and corresponding minimum in the $\nu_\mu$ survival probability, 
occurs at 24~GeV for vertically upward-going neutrinos~\cite{OscTau-Gerardo-2010}, 
so this measurement of atmospheric $\nu_e$ at higher energy is an important baseline for 
understanding the flux at energies where oscillation effects are unimportant.  

IceCube is a high-energy neutrino detector buried in the Antarctic ice.
It observes Cherenkov light from neutrino interactions.
The main detection signatures for neutrinos are long, straight tracks and approximately spherical cascades.
The former are created by neutrino-induced muons while the latter are produced by neutrino-induced electromagnetic and/or hadronic showers.
The DeepCore infill array~\cite{DCDesign-Doug-2012} to IceCube
reduces the energy threshold of
IceCube to energies as low as 10~GeV.
DeepCore's denser optical
module spacing, higher quantum efficiency photomultiplier tubes, and lower
trigger threshold, along with its deployment
in the clearest ice,
all enhance low-energy neutrino detection.

This analysis used the first IceCube data run with DeepCore, from June 2010
to May 2011 when a total of 79 IceCube strings, including
six specialized DeepCore strings, were operational. 
Each string consists of 60 digital optical modules (DOMs), equipped with a photomultipler tube~\cite{PMT-Mase-2010} 
and data acquisition electronics~\cite{DOM-Matis-2008}.
The DeepCore fiducial volume included these six strings and the seven adjacent
standard IceCube strings.
Since then, the remaining seven IceCube strings have been deployed,
including two additional DeepCore strings.

The DeepCore fiducial volume contains the 454 DOMs deployed at depths greater than 2100~m on the 13
strings of DeepCore. A dedicated DeepCore trigger~\cite{DCDesign-Doug-2012}
was run on these DOMs. It read out the full IceCube detector if photons (``hits'')
were observed in local coincidence (LC) on at least three neighboring DOMs within 2.5~$\mu$s~\cite{DOM-Matis-2008}.
The average trigger rate was 185 Hz, a factor 13 smaller than the total IceCube trigger rate.

To avoid observational bias, 10\% of the data, distributed evenly through
the year, were used to develop the
analysis and verify the detector simulation.
After application of general data quality criteria, the 281 live-days of the remaining 90\% of the data set
are used for the results presented here.

\begin{figure}[!t]
  \vspace{-5mm}
  \centering
  \includegraphics[width=3.7in]{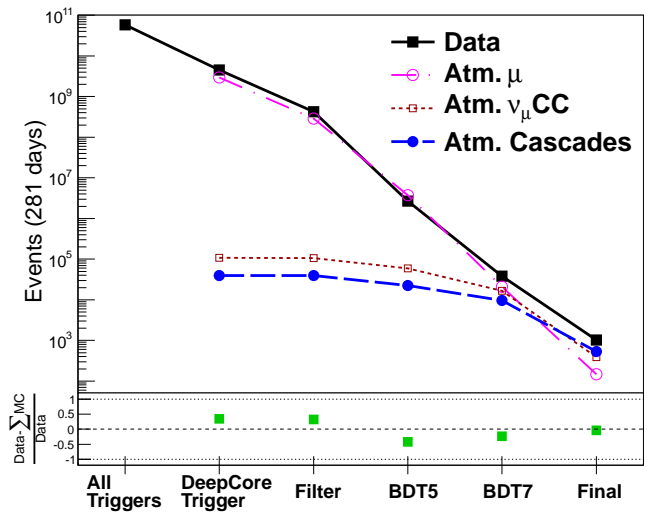}
  \caption{(Color online) The number of data events (black filled squares) passing each selection criteria is shown with the MC predictions:
    atmospheric muon~(magenta open circles), atmospheric $\nu_{\mu}$~CC~(red open squares),
    and cascade signal~(blue filled circles).
    From left to right on the horizontal axis: all triggered events, DeepCore triggered events, DeepCore filtered events,
    5-variable BDT cut (BDT5), 7-variable BDT cut(BDT7), and the final selection. Lines are to guide the eye.
  }
  \label{rate}
  \vspace{-5mm}
\end{figure}

To observe cascades, one must reject two types of
backgrounds which are more numerous than the desired signal.
The first class is ``atmospheric muons'' produced in cosmic-ray air
showers in the Earth's atmosphere, which
penetrate the 2100~m of ice above DeepCore and produce hits in the
fiducial volume.  The second class, much less common than the first,
consists of atmospheric $\nu_\mu$ also produced in air showers, which undergo CC interactions
in the ice but produce relatively low energy muons, making them
difficult to distinguish from cascades.
Simulated data are obtained from Monte Carlo (MC)
programs modeling the detector response to both of these types of
background, as well as neutrino-induced cascades.
An extensive air shower simulation~\cite{CORSIKA-Heck-1998} is used for atmospheric muons and 
a separate program~\cite{ANIS-Gazizov-2004} is for neutrinos weighted with the Honda~\cite{Honda-2007} and
the Bartol~\cite{Bartol-2004} atmospheric flux predictions.
Selection criteria are applied to the data to reduce these
backgrounds sufficiently to observe cascades. 
Fig.~\ref{rate} compares the performance of
these criteria with the predicted performance.

The initial background rejection focuses on eliminating
atmospheric muons.  Following the standard DeepCore trigger, an online veto algorithm~\cite{DCDesign-Doug-2012} 
rejects events with possible traces of an entering muon seen by the
outer strings of the IceCube detector.  A secondary trigger is
applied in software to remove spurious triggers due to dark noise, requiring at least eight hits in
the event, of which at least four must be in the fiducial volume.  In
contrast to the hardware trigger, both locally-coincident and
non-coincident signals which are correlated in time and space with the
event are included.

Next, five variables quantifying the event topology are calculated.
These are the depth of the first hit in the event, the
sphericity of the event (the ratio of the smallest eigenvalue to the sum of all eigenvalues,
the analogue of the tensor of inertia obtained by treating each hit as
a point mass), the fraction of the total number of photoelectrons which are recorded
within the first 600~ns (i.e., a measure of how fast the event develops), 
a similar fraction calculation excluding the two earliest hits assuming they may be due to noise,
and the number of hits occurring in the veto
region regardless of their time or location. These variables are used to train a boosted decision tree (BDT) \cite{BDT-Hocker-2007}.
A cut on this 5-variable BDT reduces the number of data events to 2.7$\times$10$^{6}$,
a factor of 1660 reduction from the DeepCore trigger.
The number of predicted atmospheric cascades is 2.3$\times$10$^{4}$ at this stage.

More computationally
intensive event reconstruction algorithms are run over the surviving
events, successively reconstructing the events under the hypotheses
that they are produced by atmospheric muons or by neutrino-induced
cascades. These likelihood-based reconstructions take into account the details of Cherenkov light propagation in the
ice \cite{AMANDA-Christopher-2004}. A second BDT is then trained to discriminate between
atmospheric muons and cascades, using seven variables: the
ratio of the best-fit likelihoods obtained from these reconstructions,
the depth of the first hit, the horizontal distance of the first hit
from the center of DeepCore, the fraction of the total number of photoelectrons detected
within the first 300~ns, and variables measuring whether the hit pattern
tends to drift across the detector during the event, indicative of a muon-like track through
the detector rather than a spherically expanding light pattern which is a typical
signature of a cascade. The numbers of events passing the second BDT are 
3.8$\times$10$^{4}$ (data), 2.1$\times$10$^{4}$ (atm.~$\mu$), 1.7$\times$10$^{4}$ (atm.~$\nu_\mu$~CC), 
and 9.6$\times$10$^{3}$(atm. cascades).

Following this second BDT, the atmospheric muon rate is reduced sufficiently that 
the final steps of event selection focus on reducing the background of tracklike neutrino events produced by $\nu_\mu$~CC interactions
rather than the cascades produced by $\nu_e$ or $\nu_\mu$~NC interactions. 
These are mostly $\nu_\mu$~CC events interacting within the DeepCore volume. The hadronic
cascade produced at the neutrino interaction vertex may obscure
the muon track emanating from the vertex unless the
muon has sufficient energy to travel a considerable distance and
is oriented such that it does not escape the detector
volume undetected.  To reduce this $\nu_\mu$~CC background, we apply several
additional criteria. We require that
the reconstructed neutrino interaction vertex of the event not be near the top or
bottom of the DeepCore volume, where a muon track could be missed if
it escapes into the uninstrumented ice below the detector or the
relatively dusty ice above DeepCore, and that the first hit in the
event also fall in this volume.  We further require that the best-fit
cascade likelihood be both relatively good and also better than that
obtained from the track fit, that enough DOMs were hit in the event
that the comparison between these fits should have discrimination
power, and finally we require a high (signal-like) value from the second,
7-variable BDT~\cite{Ha-Thesis-2011}.

\begin{figure}[!t]
  \centering
  \includegraphics[width=3.4in]{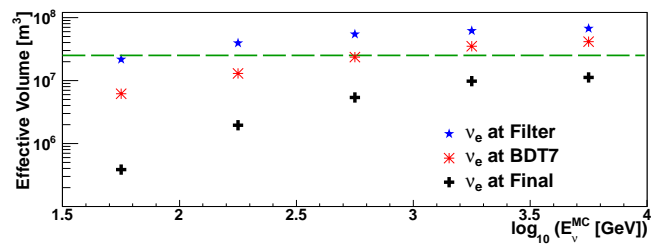}
  \vspace{-7mm}
  \caption{(Color online) Effective volumes at different cut levels as a function of MC $\nu_e$ energy. 
    The green dashed line indicates an instrumented cylinder volume with a height of 200~m and a radius of 200~m.}
  \label{veff}
  \vspace{-4mm}
\end{figure}

In contrast to the criteria used to reject atmospheric muons, this
discrimination against $\nu_\mu$~CC events leads to significant loss of neutrino cascade events,
not only due to the reduced fiducial volume but also because low
energy cascades rarely produce enough hits to satisfy the
requirements.  The steeply falling atmospheric neutrino
spectrum and the very low online trigger threshold mean that the
overall efficiency shown in Fig.~\ref{rate} is dominated by the low
energy events, but as shown in Fig.~\ref{veff}, 40\% of the
cascades within the fiducial volume pass these cuts above 1~TeV.  

In total, 1029 events pass all of these criteria. Simulations show that
about half of these events are
residual backgrounds due to atmospheric muons (14\%) and $\nu_\mu$~CC events (36\%).
The remainder are $\nu_e$ and $\nu_\mu$~NC events (inseparable from $\nu_e$).
The MC predicts that 
half of the cascades are produced by $\nu_e$ (primarily CC
interactions) and half are $\nu_\mu$~NC events.
The final event rates are shown in Table~\ref{table_results} and 
the effective volumes of $\nu_e$ simulation at different selection levels are shown in Fig.~\ref{veff}.
For atmospheric muons, the overall rejection is 4$\times$10$^8$ with respect to the total IceCube trigger rate.
These atmospheric muons mimic the cascade signature in the
fiducial region by losing most of their energy in a single
interaction; they evade the veto cuts by entering the IceCube detector
from the side on a relatively steep trajectory (typically around
60$^\circ$ in the zenith angle) in a region where the optical properties of the ice are
poorer than average \cite{AHA-Kurt-2006,Spice-Dima-2012}, so that in unfortunate cases no
Cherenkov photons are detected by the veto.  The rarity of
these events and the limited size of the simulated atmospheric muon sample
which could be produced with the available computational resources
results in a large statistical uncertainty in this background, but its
importance is clearly secondary to that of the misidentified muon
neutrino events.
\begin{table}[t]
\begin{center}
\caption{The number of events in 281 days are shown after application of all selection criteria.
  $\rm N^{obs}$ denotes the number of observed data events.
  An average of the Bartol and Honda event rates is also shown.
  The neutrino simulations have statistical uncertainties of less than 2\%
  while the atmospheric muon MC has a statistical uncertainty of 45\%.}
\begin{tabular}{c|ccc|cc|c|c}
\hline\hline
&\multicolumn{3}{|c|}{Signal} & \multicolumn{2}{|c|}{Background} &  MC &$\rm N^{obs}$\\
Type&$\nu_{e}{\rm NC}$ & $\nu_{e}{\rm CC}$ & $\nu_{\mu}{\rm NC}$ & $\nu_{\mu}{\rm CC}$ & atm.~$\mu$& Sum &\\
\hline
Bartol  & 26 & 290 & 267 & 403 & 147 & 1134 & -\\
Honda   & 19 & 227 & 245 & 368 & 147 & 1007 & - \\
Average & 23 & 259 & 256 & 385 & 147 & 1070 & - \\
Data    &  - &  -  &  -  &  -  &     -     & -& 1029\\
\hline\hline
\end{tabular}
\label{table_results}
\end{center}
  \vspace{-7mm}
\end{table}

The largest systematic uncertainty in the signal prediction comes from 
the light detection efficiency in a DOM \textit{in situ}.
The uncertainty in the absolute efficiency of the PMT is measured in a laboratory to be 8\%~\cite{PMT-Mase-2010} 
which dominates the DOM sensitivity.  Studies of efficiency in situ
with atmospheric muon data lead to an estimated 10\% uncertainty in
the DOM sensitivity when local effects in the refrozen ice are included.
Varying the efficiency by 10\% in the simulations,
the predicted atmospheric neutrino rate changes by 11\% for $\nu_\mu$ and 10\% for $\nu_e$.
The same procedure is performed for the atmospheric muons, except at an earlier stage (BDT7)
of the analysis to ensure adequate statistics and gives 30\% uncertainty.
The systematic uncertainties due to the optical properties of ice
are estimated as 8\% for atmospheric muons and 6\% (2\%) for atmospheric $\nu_\mu$ ($\nu_e$)
by comparing final level rates from simulations with two different ice models.
The optical properties of the ice are determined from measurements using calibration light source data in the DOMs~\cite{Spice-Dima-2012,AHA-Kurt-2006}.

We conservatively estimate the uncertainty in the atmospheric muon rate
due to uncertainties in cosmic ray composition by comparing our baseline simulation,
based on spectra of individual elements~\cite{PolyGonato-Hoerandel-2003} with a proton-only composition model.
The comparison is made at the BDT7 stage to ensure sufficient statistics,
and shows a rate variation of 25\%. Additionally, a 20\% uncertainty for the cosmic ray flux normalization and 6\% for the seasonal rate variation are included. 
These are summed in quadrature and give a total of 33\% cosmic-ray flux uncertainty.
Though large, this cosmic-ray flux uncertainty is smaller than the statistical uncertainty
in the atmospheric muon rate due to the limited MC sample, so we use this estimate as the systematic uncertainty.
The systematic uncertainty for neutrino-nucleon cross sections is estimated to be 6\%.
The atmospheric $\nu_\mu$ flux uncertainties of 9\% are obtained
by comparing the final event rates with the Honda and Bartol flux predictions.
Neutrino oscillations have a very small effect in this sample (1.8\% for $\nu_{\mu}$ and 0.1\% for $\nu_{e}$).
The $\nu_\tau$ contribution is estimated to be less than 1\% of the data sample
assuming standard oscillation parameters~\cite{Osc-GonzalezGarcia-2010}.
The total systematic uncertainties are 14\% (11\%) for atmospheric $\nu_\mu$ ($\nu_e$) and 45\% for atmospheric muons, 
as shown in Table~\ref{systematic}.
\begin{table}[t]
\begin{center}
\caption{Systematic uncertainties.}
\begin{tabular}{c|c|c|c}
\hline\hline
Source of uncertainties & atm.~$\mu$ & atm.~$\nu_{\mu}$ & atm.~$\nu_{e}$  \\
\hline
Ice properties & 8\% & 6\% & 2\%  \\
DOM efficiency & 30\%  & 11\% & 10\%\\
Cosmic-ray flux & 33\% & - & - \\
$\nu$-nucleon cross section & - & 6\% & 6\%\\
\hline
Sum  & 45\% & 14\% & 11\%\\
\hline\hline
\end{tabular}
\label{systematic}
\end{center}
  \vspace{-5mm}
\end{table}

The atmospheric muon background and the atmospheric $\nu_\mu$~CC background
are subtracted from the data. The latter contribution is estimated by averaging
the Bartol and Honda atmospheric neutrino predictions.
Half the difference is included in the systematic uncertainties.
We observe an excess of cascade events,
\begin{eqnarray}  \label{eqn:ccR}
  N_{\rm cascade} &=&
  \begin{array}{l l}
    496 \pm 66 (\text{stat.}) \pm 88 (\text{syst.})
  \end{array}\nonumber,
\end{eqnarray}
where the total statistical uncertainty includes statistical uncertainties of the two subtracted background components,
and the total systematic uncertainty is a sum in quadrature
of the $\nu_\mu$~CC systematic uncertainties and the atmospheric muon systematic uncertainties.
Since part of the systematic uncertainties does not come from the final level comparison, 
we conservatively do not consider the correlations among the systematic uncertainties.
The cascade signal has a significance of 4.5~$\sigma$.
We estimate based on simulations that 240 $\pm$ 66(stat.) $\pm$ 109(syst.) of the cascades are produced by $\nu_e$.
The data are in good agreement with the Honda model, and slightly below (though still consistent with) the Bartol model which predicts 127 more neutrino events in total.

The lower rate prediction from the Honda model, especially in $\nu_e$, is due to the different treatment
of kaon production in the atmosphere~\cite{IntHonda-2007},
and is shown in Table~\ref{table_results}.
Both Honda and Bartol estimate roughly 15\% uncertainties in
the atmospheric $\nu_e$ flux at 100~GeV rising to
25\% at 1~TeV~\cite{Honda-2007, Bartol-2004, BartolError-2006}.
\begin{figure}[!t]
  \vspace{-3mm}
  \centering
  \includegraphics[width=3.7in]{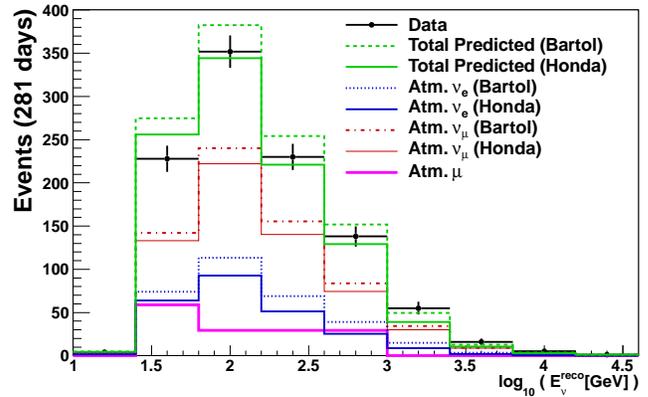}
  \vspace{-7mm}
  \caption{(Color online) The event rate as a function of the reconstructed cascade energy.
    The sum of all MC expectations (green) is consistent with 281 days of data rate.
    The dotted lines show the Bartol prediction while the solid lines indicate Honda predictions for
    the atmospheric neutrinos.
    Systematic uncertainties (Table~\ref{systematic}) are omitted for clarity.}
  \label{energy}
  \vspace{-5mm}
\end{figure}

\begin{figure}[!t]
  \centering
  \includegraphics[width=3.4in]{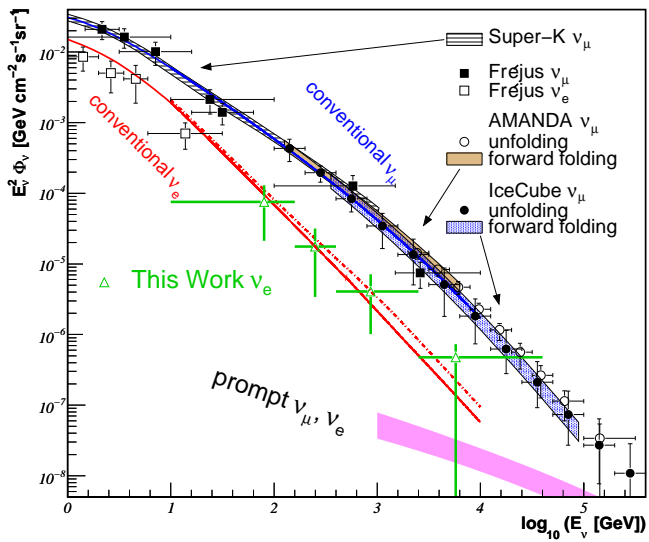}
  \vspace{-7mm}
  \caption{(Color online) The electron neutrino spectrum (green open triangles).
    The conventional $\nu_e$ (red line) and $\nu_\mu$ (blue line) from Honda, $\nu_e$ (red dotted line) from Bartol and
    charm-induced neutrinos (magenta band)~\cite{Charm-Enberg-2008} are shown.
    Previous measurements from Super-K~\cite{SKflux-GonzalezGarcia-2006}, Fr\'ejus~\cite{Frejus-Daum-1995}, 
    AMANDA~\cite{AMANDA-Kelley-2009, AMANDA-Julia-2010} and IceCube~\cite{Diffuse-Warren-2011, Diffuse-Sean-2011} are also shown.
  }
  \label{nueflux}
  \vspace{-5mm}
\end{figure}

Likelihood reconstructions are performed on every event in the final sample,
simultaneously fitting a cascade hypothesis for deposited energy and vertex position and time.
A vertex resolution of 9~m and an energy resolution of 0.12 in $\rm log_{10} (E/GeV)$ are obtained.
The absolute energy scale uncertainty is found to be 0.1 in $\rm log_{10} (E/GeV)$.
Using the energy reconstruction in Fig.~\ref{energy} (rebinned to get sufficient statistics and 
reasonable uncertainties in each bin), we subtract the atmospheric muon and 
the atmospheric $\nu_\mu$~CC and $\nu_\mu$~NC to estimate the $\nu_e$ excess.
The $\nu_e$ excess is converted into flux by normalizing to the expected number of events 
from an average of the Bartol and Honda fluxes. 
In each bin, the horizontal bar indicates the bin width. 
The marker placement shows the average reconstructed energy 
of the contributing events. 
The vertical error bars include the statistical and 
systematic uncertainties (see Fig.~\ref{nueflux} and Table~\ref{fluxtable}).

\begin{table}[t]
  \begin{center}
\caption{The $\rm E^2_\nu\Phi_\nu$ flux. $\rm E_\nu$ is in GeV.}
\begin{tabular}{c|c|cc}
\hline
$\rm log_{10}E_\nu^{min} - log_{10}E_\nu^{max} $& $\rm \langle E_{\nu} \rangle$ &$\rm E^2_\nu\Phi_\nu$($\rm GeV~cm^{-2}s^{-1}sr^{-1}$)& \\
\hline
1.0 $-$ 2.2   & 80 & $(7.5\pm5.4) \times 10^{-5}$     &      \\
2.2 $-$ 2.6   & 251 & $(1.8\pm1.4) \times 10^{-5}$     &       \\
2.6 $-$ 3.4   & 865 & $(4.1\pm3.1) \times 10^{-6}$     &       \\
3.4 $-$ 4.6   & 5753 & $4.8^{+2.6}_{-4.8} \times 10^{-7}$     &       \\
\hline
\end{tabular}
\label{fluxtable}
\end{center}
  \vspace{-7mm}
\end{table}

In conclusion, we have observed
atmospheric neutrino-induced cascades, produced by $\nu_{e}$~CC interactions and NC interactions of all flavors in IceCube.
The atmospheric $\nu_e$ flux in the energy range between 80~GeV and 6~TeV
is consistent with current models of the atmospheric neutrino flux.  
More sophisticated event reconstruction algorithms now in development,
combined with the additional information from the final two DeepCore
strings deployed in late 2010, should provide substantially improved
discrimination against the $\nu_\mu$~CC background.  This will provide both a
more precise measurement of the electron neutrino flux and a reduced
energy threshold, enabling observation of oscillation phenomena in the
cascade channel.  
\begin{acknowledgments}
We acknowledge the support from the following agencies:
U.S. National Science Foundation-Office of Polar Programs,
U.S. National Science Foundation-Physics Division,
University of Wisconsin Alumni Research Foundation,
the Grid Laboratory Of Wisconsin (GLOW) grid infrastructure at the University of Wisconsin - Madison, the Open Science Grid (OSG) grid infrastructure;
U.S. Department of Energy, and National Energy Research Scientific Computing Center,
the Louisiana Optical Network Initiative (LONI) grid computing resources;
National Science and Engineering Research Council of Canada;
Swedish Research Council,
Swedish Polar Research Secretariat,
Swedish National Infrastructure for Computing (SNIC),
and Knut and Alice Wallenberg Foundation, Sweden;
German Ministry for Education and Research (BMBF),
Deutsche Forschungsgemeinschaft (DFG),
Helmholtz Alliance for Astroparticle Physics (HAP),
Research Department of Plasmas with Complex Interactions (Bochum), Germany;
Fund for Scientific Research (FNRS-FWO),
FWO Odysseus programme,
Flanders Institute to encourage scientific and technological research in industry (IWT),
Belgian Federal Science Policy Office (Belspo);
University of Oxford, United Kingdom;
Marsden Fund, New Zealand;
Australian Research Council;
Japan Society for Promotion of Science (JSPS);
the Swiss National Science Foundation (SNSF), Switzerland.
\end{acknowledgments}

\begin{thebibliography}{33}%
\makeatletter
\providecommand \@ifxundefined [1]{%
 \@ifx{#1\undefined}
}%
\providecommand \@ifnum [1]{%
 \ifnum #1\expandafter \@firstoftwo
 \else \expandafter \@secondoftwo
 \fi
}%
\providecommand \@ifx [1]{%
 \ifx #1\expandafter \@firstoftwo
 \else \expandafter \@secondoftwo
 \fi
}%
\providecommand \natexlab [1]{#1}%
\providecommand \enquote  [1]{``#1''}%
\providecommand \bibnamefont  [1]{#1}%
\providecommand \bibfnamefont [1]{#1}%
\providecommand \citenamefont [1]{#1}%
\providecommand \href@noop [0]{\@secondoftwo}%
\providecommand \href [0]{\begingroup \@sanitize@url \@href}%
\providecommand \@href[1]{\@@startlink{#1}\@@href}%
\providecommand \@@href[1]{\endgroup#1\@@endlink}%
\providecommand \@sanitize@url [0]{\catcode `\\12\catcode `\$12\catcode
  `\&12\catcode `\#12\catcode `\^12\catcode `\_12\catcode `\%12\relax}%
\providecommand \@@startlink[1]{}%
\providecommand \@@endlink[0]{}%
\providecommand \url  [0]{\begingroup\@sanitize@url \@url }%
\providecommand \@url [1]{\endgroup\@href {#1}{\urlprefix }}%
\providecommand \urlprefix  [0]{URL }%
\providecommand \Eprint [0]{\href }%
\providecommand \doibase [0]{http://dx.doi.org/}%
\providecommand \selectlanguage [0]{\@gobble}%
\providecommand \bibinfo  [0]{\@secondoftwo}%
\providecommand \bibfield  [0]{\@secondoftwo}%
\providecommand \translation [1]{[#1]}%
\providecommand \BibitemOpen [0]{}%
\providecommand \bibitemStop [0]{}%
\providecommand \bibitemNoStop [0]{.\EOS\space}%
\providecommand \EOS [0]{\spacefactor3000\relax}%
\providecommand \BibitemShut  [1]{\csname bibitem#1\endcsname}%
\let\auto@bib@innerbib\@empty
\bibitem [{\citenamefont {Abbasi}\ \emph
  {et~al.}(2011{\natexlab{a}})\citenamefont {Abbasi} \emph
  {et~al.}}]{Diffuse-Warren-2011}%
  \BibitemOpen
  \bibfield  {author} {\bibinfo {author} {\bibfnamefont {R.}~\bibnamefont
  {Abbasi}} \emph {et~al.} (\bibinfo {collaboration} {IceCube Collaboration}),\
  }\href {\doibase 10.1103/PhysRevD.83.012001} {\bibfield  {journal} {\bibinfo
  {journal} {Phys. Rev. D}\ }\textbf {\bibinfo {volume} {83}},\ \bibinfo
  {pages} {012001} (\bibinfo {year} {2011}{\natexlab{a}})}\BibitemShut
  {NoStop}%
\bibitem [{\citenamefont {Becker-Szendy}\ \emph {et~al.}(1992)\citenamefont
  {Becker-Szendy} \emph {et~al.}}]{IMB-Becker-1992}%
  \BibitemOpen
  \bibfield  {author} {\bibinfo {author} {\bibfnamefont {R.}~\bibnamefont
  {Becker-Szendy}} \emph {et~al.} (\bibinfo {collaboration} {IMB-3
  Collaboration}),\ }\href {\doibase 10.1103/PhysRevD.46.3720} {\bibfield
  {journal} {\bibinfo  {journal} {Phys. Rev. D}\ }\textbf {\bibinfo {volume}
  {46}},\ \bibinfo {pages} {3720} (\bibinfo {year} {1992})}\BibitemShut
  {NoStop}%
\bibitem [{\citenamefont {Wendell}\ \emph {et~al.}(2010)\citenamefont {Wendell}
  \emph {et~al.}}]{SK-Wendell-2010}%
  \BibitemOpen
  \bibfield  {author} {\bibinfo {author} {\bibfnamefont {R.}~\bibnamefont
  {Wendell}} \emph {et~al.} (\bibinfo {collaboration} {Super-Kamiokande
  Collaboration}),\ }\href {\doibase 10.1103/PhysRevD.81.092004} {\bibfield
  {journal} {\bibinfo  {journal} {Phys. Rev. D}\ }\textbf {\bibinfo {volume}
  {81}},\ \bibinfo {pages} {092004} (\bibinfo {year} {2010})}\BibitemShut
  {NoStop}%
\bibitem [{\citenamefont {Daum}\ \emph {et~al.}(1995)\citenamefont {Daum} \emph
  {et~al.}}]{Frejus-Daum-1995}%
  \BibitemOpen
  \bibfield  {author} {\bibinfo {author} {\bibfnamefont {K.}~\bibnamefont
  {Daum}} \emph {et~al.} (\bibinfo {collaboration} {Fr\'ejus Collaboration}),\
  }\href@noop {} {\bibfield  {journal} {\bibinfo  {journal} {Z. Phys. C}\
  }\textbf {\bibinfo {volume} {66}},\ \bibinfo {pages} {417} (\bibinfo {year}
  {1995})}\BibitemShut {NoStop}%
\bibitem [{\citenamefont {Aglietta}\ \emph {et~al.}(1989)\citenamefont
  {Aglietta} \emph {et~al.}}]{NUSEX-Aglietta-1988}%
  \BibitemOpen
  \bibfield  {author} {\bibinfo {author} {\bibfnamefont {M.}~\bibnamefont
  {Aglietta}} \emph {et~al.} (\bibinfo {collaboration} {NUSEX Collaboration}),\
  }\href {\doibase 10.1209/0295-5075/8/7/005} {\bibfield  {journal} {\bibinfo
  {journal} {Europhys. Lett.}\ }\textbf {\bibinfo {volume} {8}},\ \bibinfo
  {pages} {611} (\bibinfo {year} {1989})}\BibitemShut {NoStop}%
\bibitem [{\citenamefont {Sanchez}\ \emph {et~al.}(2003)\citenamefont {Sanchez}
  \emph {et~al.}}]{Soudan2-Sanchez-2003}%
  \BibitemOpen
  \bibfield  {author} {\bibinfo {author} {\bibfnamefont {M.}~\bibnamefont
  {Sanchez}} \emph {et~al.} (\bibinfo {collaboration} {Soudan-2
  Collaboration}),\ }\href {\doibase 10.1103/PhysRevD.68.113004} {\bibfield
  {journal} {\bibinfo  {journal} {Phys. Rev. D}\ }\textbf {\bibinfo {volume}
  {68}},\ \bibinfo {pages} {113004} (\bibinfo {year} {2003})}\BibitemShut
  {NoStop}%
\bibitem [{\citenamefont {Ackermann}\ \emph {et~al.}(2004)\citenamefont
  {Ackermann} \emph {et~al.}}]{Ackermann:2004zw}%
  \BibitemOpen
  \bibfield  {author} {\bibinfo {author} {\bibfnamefont {M.}~\bibnamefont
  {Ackermann}} \emph {et~al.} (\bibinfo {collaboration} {AMANDA
  Collaboration}),\ }\href {\doibase 10.1016/j.astropartphys.2004.06.003}
  {\bibfield  {journal} {\bibinfo  {journal} {Astropart. Phys.}\ }\textbf
  {\bibinfo {volume} {22}},\ \bibinfo {pages} {127} (\bibinfo {year}
  {2004})}\BibitemShut {NoStop}%
\bibitem [{\citenamefont {Abbasi}\ \emph
  {et~al.}(2011{\natexlab{b}})\citenamefont {Abbasi} \emph
  {et~al.}}]{Cascade-Mike-2011}%
  \BibitemOpen
  \bibfield  {author} {\bibinfo {author} {\bibfnamefont {R.}~\bibnamefont
  {Abbasi}} \emph {et~al.} (\bibinfo {collaboration} {IceCube Collaboration}),\
  }\href@noop {} {\bibfield  {journal} {\bibinfo  {journal} {Astropart. Phys.}\
  }\textbf {\bibinfo {volume} {34}},\ \bibinfo {pages} {420} (\bibinfo {year}
  {2011}{\natexlab{b}})}\BibitemShut {NoStop}%
\bibitem [{\citenamefont {Abbasi}\ \emph
  {et~al.}(2011{\natexlab{c}})\citenamefont {Abbasi} \emph
  {et~al.}}]{Cascade-Joanna-2011}%
  \BibitemOpen
  \bibfield  {author} {\bibinfo {author} {\bibfnamefont {R.}~\bibnamefont
  {Abbasi}} \emph {et~al.} (\bibinfo {collaboration} {IceCube Collaboration}),\
  }\href {\doibase 10.1103/PhysRevD.84.072001} {\bibfield  {journal} {\bibinfo
  {journal} {Phys. Rev. D}\ }\textbf {\bibinfo {volume} {84}},\ \bibinfo
  {pages} {072001} (\bibinfo {year} {2011}{\natexlab{c}})}\BibitemShut
  {NoStop}%
\bibitem [{\citenamefont {Abbasi}\ \emph
  {et~al.}(2011{\natexlab{d}})\citenamefont {Abbasi} \emph
  {et~al.}}]{Cascade-ICRC-2011}%
  \BibitemOpen
  \bibfield  {author} {\bibinfo {author} {\bibfnamefont {R.}~\bibnamefont
  {Abbasi}} \emph {et~al.} (\bibinfo {collaboration} {IceCube Collaboration}),\
  }in\ \href@noop {} {\emph {\bibinfo {booktitle} {Proc. of 32nd ICRC, Beijing,
  China}}}\ (\bibinfo {year} {2011})\ \Eprint {http://arxiv.org/abs/1111.2736}
  {arXiv:1111.2736} \BibitemShut {NoStop}%
\bibitem [{\citenamefont {Honda}\ \emph {et~al.}(2007)\citenamefont {Honda},
  \citenamefont {Kajita}, \citenamefont {Kasahara}, \citenamefont
  {Midorikawa},\ and\ \citenamefont {Sanuki}}]{Honda-2007}%
  \BibitemOpen
  \bibfield  {author} {\bibinfo {author} {\bibfnamefont {M.}~\bibnamefont
  {Honda}}, \bibinfo {author} {\bibfnamefont {T.}~\bibnamefont {Kajita}},
  \bibinfo {author} {\bibfnamefont {K.}~\bibnamefont {Kasahara}}, \bibinfo
  {author} {\bibfnamefont {S.}~\bibnamefont {Midorikawa}}, \ and\ \bibinfo
  {author} {\bibfnamefont {T.}~\bibnamefont {Sanuki}},\ }\href {\doibase
  10.1103/PhysRevD.75.043006} {\bibfield  {journal} {\bibinfo  {journal} {Phys.
  Rev. D}\ }\textbf {\bibinfo {volume} {75}},\ \bibinfo {pages} {043006}
  (\bibinfo {year} {2007})}\BibitemShut {NoStop}%
\bibitem [{\citenamefont {Barr}\ \emph {et~al.}(2006)\citenamefont {Barr},
  \citenamefont {Robbins}, \citenamefont {Gaisser},\ and\ \citenamefont
  {Stanev}}]{BartolError-2006}%
  \BibitemOpen
  \bibfield  {author} {\bibinfo {author} {\bibfnamefont {G.~D.}\ \bibnamefont
  {Barr}}, \bibinfo {author} {\bibfnamefont {S.}~\bibnamefont {Robbins}},
  \bibinfo {author} {\bibfnamefont {T.~K.}\ \bibnamefont {Gaisser}}, \ and\
  \bibinfo {author} {\bibfnamefont {T.}~\bibnamefont {Stanev}},\ }\href
  {\doibase 10.1103/PhysRevD.74.094009} {\bibfield  {journal} {\bibinfo
  {journal} {Phys. Rev. D}\ }\textbf {\bibinfo {volume} {74}},\ \bibinfo
  {pages} {094009} (\bibinfo {year} {2006})}\BibitemShut {NoStop}%
\bibitem [{\citenamefont {Koskinen}(2011)}]{DC-Koskinen-2011}%
  \BibitemOpen
  \bibfield  {author} {\bibinfo {author} {\bibfnamefont {D.~J.}\ \bibnamefont
  {Koskinen}},\ }\href {\doibase 10.1142/S021773231103725X} {\bibfield
  {journal} {\bibinfo  {journal} {Mod. Phys. Lett.}\ }\textbf {\bibinfo
  {volume} {A26}},\ \bibinfo {pages} {2899} (\bibinfo {year}
  {2011})}\BibitemShut {NoStop}%
\bibitem [{\citenamefont {Giordano}\ \emph {et~al.}(2010)\citenamefont
  {Giordano}, \citenamefont {Mena},\ and\ \citenamefont
  {Mocioiu}}]{OscTau-Gerardo-2010}%
  \BibitemOpen
  \bibfield  {author} {\bibinfo {author} {\bibfnamefont {G.}~\bibnamefont
  {Giordano}}, \bibinfo {author} {\bibfnamefont {O.}~\bibnamefont {Mena}}, \
  and\ \bibinfo {author} {\bibfnamefont {I.}~\bibnamefont {Mocioiu}},\ }\href
  {\doibase 10.1103/PhysRevD.81.113008} {\bibfield  {journal} {\bibinfo
  {journal} {Phys. Rev. D}\ }\textbf {\bibinfo {volume} {81}},\ \bibinfo
  {pages} {113008} (\bibinfo {year} {2010})}\BibitemShut {NoStop}%
\bibitem [{\citenamefont {Abbasi}\ \emph {et~al.}(2012)\citenamefont {Abbasi}
  \emph {et~al.}}]{DCDesign-Doug-2012}%
  \BibitemOpen
  \bibfield  {author} {\bibinfo {author} {\bibfnamefont {R.}~\bibnamefont
  {Abbasi}} \emph {et~al.} (\bibinfo {collaboration} {IceCube Collaboration}),\
  }\href@noop {} {\bibfield  {journal} {\bibinfo  {journal} {Astropart. Phys.}\
  }\textbf {\bibinfo {volume} {35}},\ \bibinfo {pages} {615} (\bibinfo {year}
  {2012})}\BibitemShut {NoStop}%
\bibitem [{\citenamefont {Abbasi}\ \emph
  {et~al.}(2010{\natexlab{a}})\citenamefont {Abbasi} \emph
  {et~al.}}]{PMT-Mase-2010}%
  \BibitemOpen
  \bibfield  {author} {\bibinfo {author} {\bibfnamefont {R.}~\bibnamefont
  {Abbasi}} \emph {et~al.} (\bibinfo {collaboration} {IceCube Collaboration}),\
  }\href {\doibase 10.1016/j.nima.2010.03.102} {\bibfield  {journal} {\bibinfo
  {journal} {Nucl. Instrum. Meth.}\ }\textbf {\bibinfo {volume} {A618}},\
  \bibinfo {pages} {139} (\bibinfo {year} {2010}{\natexlab{a}})}\BibitemShut
  {NoStop}%
\bibitem [{\citenamefont {Abbasi}\ \emph
  {et~al.}(2009{\natexlab{a}})\citenamefont {Abbasi} \emph
  {et~al.}}]{DOM-Matis-2008}%
  \BibitemOpen
  \bibfield  {author} {\bibinfo {author} {\bibfnamefont {R.}~\bibnamefont
  {Abbasi}} \emph {et~al.} (\bibinfo {collaboration} {IceCube Collaboration}),\
  }\href {\doibase 10.1016/j.nima.2009.01.001} {\bibfield  {journal} {\bibinfo
  {journal} {Nucl. Instrum. Meth.}\ }\textbf {\bibinfo {volume} {A601}},\
  \bibinfo {pages} {294} (\bibinfo {year} {2009}{\natexlab{a}})}\BibitemShut
  {NoStop}%
\bibitem [{\citenamefont {Heck}\ \emph {et~al.}(1998)\citenamefont {Heck} \emph
  {et~al.}}]{CORSIKA-Heck-1998}%
  \BibitemOpen
  \bibfield  {author} {\bibinfo {author} {\bibfnamefont {D.}~\bibnamefont
  {Heck}} \emph {et~al.},\ }\href@noop {} {\bibfield  {journal} {\bibinfo
  {journal} {Tech. Rep. FZKA}\ }\textbf {\bibinfo {volume} {6019}},\ \bibinfo
  {pages} {1} (\bibinfo {year} {1998})}\BibitemShut {NoStop}%
\bibitem [{\citenamefont {Gazizov}\ and\ \citenamefont
  {Kowalski}(2005)}]{ANIS-Gazizov-2004}%
  \BibitemOpen
  \bibfield  {author} {\bibinfo {author} {\bibfnamefont {A.}~\bibnamefont
  {Gazizov}}\ and\ \bibinfo {author} {\bibfnamefont {M.~P.}\ \bibnamefont
  {Kowalski}},\ }\href {\doibase 10.1016/j.cpc.2005.03.113} {\bibfield
  {journal} {\bibinfo  {journal} {Comput. Phys. Commun.}\ }\textbf {\bibinfo
  {volume} {172}},\ \bibinfo {pages} {203} (\bibinfo {year}
  {2005})}\BibitemShut {NoStop}%
\bibitem [{\citenamefont {Barr}\ \emph {et~al.}(2004)\citenamefont {Barr},
  \citenamefont {Gaisser}, \citenamefont {Lipari}, \citenamefont {Robbins},\
  and\ \citenamefont {Stanev}}]{Bartol-2004}%
  \BibitemOpen
  \bibfield  {author} {\bibinfo {author} {\bibfnamefont {G.~D.}\ \bibnamefont
  {Barr}}, \bibinfo {author} {\bibfnamefont {T.~K.}\ \bibnamefont {Gaisser}},
  \bibinfo {author} {\bibfnamefont {P.}~\bibnamefont {Lipari}}, \bibinfo
  {author} {\bibfnamefont {S.}~\bibnamefont {Robbins}}, \ and\ \bibinfo
  {author} {\bibfnamefont {T.}~\bibnamefont {Stanev}},\ }\href {\doibase
  10.1103/PhysRevD.70.023006} {\bibfield  {journal} {\bibinfo  {journal} {Phys.
  Rev. D}\ }\textbf {\bibinfo {volume} {70}},\ \bibinfo {pages} {023006}
  (\bibinfo {year} {2004})}\BibitemShut {NoStop}%
\bibitem [{\citenamefont {Hocker}\ \emph {et~al.}(2007)\citenamefont {Hocker},
  \citenamefont {Stelzer}, \citenamefont {Tegenfeldt}, \citenamefont {Voss},
  \citenamefont {Voss} \emph {et~al.}}]{BDT-Hocker-2007}%
  \BibitemOpen
  \bibfield  {author} {\bibinfo {author} {\bibfnamefont {A.}~\bibnamefont
  {Hocker}}, \bibinfo {author} {\bibfnamefont {J.}~\bibnamefont {Stelzer}},
  \bibinfo {author} {\bibfnamefont {F.}~\bibnamefont {Tegenfeldt}}, \bibinfo
  {author} {\bibfnamefont {H.}~\bibnamefont {Voss}}, \bibinfo {author}
  {\bibfnamefont {K.}~\bibnamefont {Voss}},  \emph {et~al.},\ }\href@noop {}
  {\bibfield  {journal} {\bibinfo  {journal} {PoS}\ }\textbf {\bibinfo {volume}
  {ACAT}},\ \bibinfo {pages} {040} (\bibinfo {year} {2007})}\BibitemShut
  {NoStop}%
\bibitem [{\citenamefont {Ahrens}\ \emph {et~al.}(2004)\citenamefont {Ahrens}
  \emph {et~al.}}]{AMANDA-Christopher-2004}%
  \BibitemOpen
  \bibfield  {author} {\bibinfo {author} {\bibfnamefont {J.}~\bibnamefont
  {Ahrens}} \emph {et~al.} (\bibinfo {collaboration} {AMANDA Collaboration}),\
  }\href {\doibase 10.1016/j.nima.2004.01.065} {\bibfield  {journal} {\bibinfo
  {journal} {Nucl. Instrum. Meth.}\ }\textbf {\bibinfo {volume} {A524}},\
  \bibinfo {pages} {169} (\bibinfo {year} {2004})}\BibitemShut {NoStop}%
\bibitem [{\citenamefont {Ha}(2011)}]{Ha-Thesis-2011}%
  \BibitemOpen
  \bibfield  {author} {\bibinfo {author} {\bibfnamefont {C.}~\bibnamefont
  {Ha}},\ }\href@noop {} {\bibfield  {journal} {\bibinfo  {journal}
  {Ph.D.~dissertation, Pennsylvania State University
  https://etda.libraries.psu.edu/paper/12119}\ } (\bibinfo {year}
  {2011})}\BibitemShut {NoStop}%
\bibitem [{\citenamefont {{Ackermann}}\ \emph {et~al.}(2006)\citenamefont
  {{Ackermann}} \emph {et~al.}}]{AHA-Kurt-2006}%
  \BibitemOpen
  \bibfield  {author} {\bibinfo {author} {\bibfnamefont {M.}~\bibnamefont
  {{Ackermann}}} \emph {et~al.} (\bibinfo {collaboration} {The IceCube
  Collaboration}),\ }\href@noop {} {\bibfield  {journal} {\bibinfo  {journal}
  {J. Geophys. Res.}\ }\textbf {\bibinfo {volume} {111}},\ \bibinfo {pages}
  {D13203} (\bibinfo {year} {2006})}\BibitemShut {NoStop}%
\bibitem [{\citenamefont {Aartsen}\ \emph {et~al.}(2013)\citenamefont {Aartsen}
  \emph {et~al.}}]{Spice-Dima-2012}%
  \BibitemOpen
  \bibfield  {author} {\bibinfo {author} {\bibfnamefont {M.}~\bibnamefont
  {Aartsen}} \emph {et~al.} (\bibinfo {collaboration} {IceCube
  Collaboration}),\ }\href {\doibase 10.1016/j.nima.2013.01.054} {\bibfield
  {journal} {\bibinfo  {journal} {Nucl. Instrum. Meth.}\ }\textbf {\bibinfo
  {volume} {A711}},\ \bibinfo {pages} {73} (\bibinfo {year}
  {2013})}\BibitemShut {NoStop}%
\bibitem [{\citenamefont {Hoerandel}(2003)}]{PolyGonato-Hoerandel-2003}%
  \BibitemOpen
  \bibfield  {author} {\bibinfo {author} {\bibfnamefont {J.~R.}\ \bibnamefont
  {Hoerandel}},\ }\href {\doibase 10.1016/S0927-6505(02)00198-6} {\bibfield
  {journal} {\bibinfo  {journal} {Astropart. Phys.}\ }\textbf {\bibinfo
  {volume} {19}},\ \bibinfo {pages} {193} (\bibinfo {year} {2003})}\BibitemShut
  {NoStop}%
\bibitem [{\citenamefont {Gonzalez-Garcia}\ \emph {et~al.}(2010)\citenamefont
  {Gonzalez-Garcia}, \citenamefont {Maltoni},\ and\ \citenamefont
  {Salvado}}]{Osc-GonzalezGarcia-2010}%
  \BibitemOpen
  \bibfield  {author} {\bibinfo {author} {\bibfnamefont {M.}~\bibnamefont
  {Gonzalez-Garcia}}, \bibinfo {author} {\bibfnamefont {M.}~\bibnamefont
  {Maltoni}}, \ and\ \bibinfo {author} {\bibfnamefont {J.}~\bibnamefont
  {Salvado}},\ }\href {\doibase 10.1007/JHEP04(2010)056} {\bibfield  {journal}
  {\bibinfo  {journal} {JHEP}\ }\textbf {\bibinfo {volume} {1004}},\ \bibinfo
  {pages} {056} (\bibinfo {year} {2010})}\BibitemShut {NoStop}%
\bibitem [{\citenamefont {Sanuki}\ \emph {et~al.}(2007)\citenamefont {Sanuki},
  \citenamefont {Honda}, \citenamefont {Kajita}, \citenamefont {Kasahara},\
  and\ \citenamefont {Midorikawa}}]{IntHonda-2007}%
  \BibitemOpen
  \bibfield  {author} {\bibinfo {author} {\bibfnamefont {T.}~\bibnamefont
  {Sanuki}}, \bibinfo {author} {\bibfnamefont {M.}~\bibnamefont {Honda}},
  \bibinfo {author} {\bibfnamefont {T.}~\bibnamefont {Kajita}}, \bibinfo
  {author} {\bibfnamefont {K.}~\bibnamefont {Kasahara}}, \ and\ \bibinfo
  {author} {\bibfnamefont {S.}~\bibnamefont {Midorikawa}},\ }\href {\doibase
  10.1103/PhysRevD.75.043005} {\bibfield  {journal} {\bibinfo  {journal} {Phys.
  Rev. D}\ }\textbf {\bibinfo {volume} {75}},\ \bibinfo {pages} {043005}
  (\bibinfo {year} {2007})}\BibitemShut {NoStop}%
\bibitem [{\citenamefont {Enberg}\ \emph {et~al.}(2008)\citenamefont {Enberg},
  \citenamefont {Reno},\ and\ \citenamefont {Sarcevic}}]{Charm-Enberg-2008}%
  \BibitemOpen
  \bibfield  {author} {\bibinfo {author} {\bibfnamefont {R.}~\bibnamefont
  {Enberg}}, \bibinfo {author} {\bibfnamefont {M.~H.}\ \bibnamefont {Reno}}, \
  and\ \bibinfo {author} {\bibfnamefont {I.}~\bibnamefont {Sarcevic}},\ }\href
  {\doibase 10.1103/PhysRevD.78.043005} {\bibfield  {journal} {\bibinfo
  {journal} {Phys. Rev. D}\ }\textbf {\bibinfo {volume} {78}},\ \bibinfo
  {pages} {043005} (\bibinfo {year} {2008})}\BibitemShut {NoStop}%
\bibitem [{\citenamefont {Gonzalez-Garcia}\ \emph {et~al.}(2006)\citenamefont
  {Gonzalez-Garcia}, \citenamefont {Maltoni},\ and\ \citenamefont
  {Rojo}}]{SKflux-GonzalezGarcia-2006}%
  \BibitemOpen
  \bibfield  {author} {\bibinfo {author} {\bibfnamefont {M.}~\bibnamefont
  {Gonzalez-Garcia}}, \bibinfo {author} {\bibfnamefont {M.}~\bibnamefont
  {Maltoni}}, \ and\ \bibinfo {author} {\bibfnamefont {J.}~\bibnamefont
  {Rojo}},\ }\href {\doibase 10.1088/1126-6708/2006/10/075} {\bibfield
  {journal} {\bibinfo  {journal} {JHEP}\ }\textbf {\bibinfo {volume} {0610}},\
  \bibinfo {pages} {075} (\bibinfo {year} {2006})}\BibitemShut {NoStop}%
\bibitem [{\citenamefont {Abbasi}\ \emph
  {et~al.}(2009{\natexlab{b}})\citenamefont {Abbasi} \emph
  {et~al.}}]{AMANDA-Kelley-2009}%
  \BibitemOpen
  \bibfield  {author} {\bibinfo {author} {\bibfnamefont {R.}~\bibnamefont
  {Abbasi}} \emph {et~al.} (\bibinfo {collaboration} {IceCube Collaboration}),\
  }\href {\doibase 10.1103/PhysRevD.79.102005} {\bibfield  {journal} {\bibinfo
  {journal} {Phys. Rev. D}\ }\textbf {\bibinfo {volume} {79}},\ \bibinfo
  {pages} {102005} (\bibinfo {year} {2009}{\natexlab{b}})}\BibitemShut
  {NoStop}%
\bibitem [{\citenamefont {Abbasi}\ \emph
  {et~al.}(2010{\natexlab{b}})\citenamefont {Abbasi} \emph
  {et~al.}}]{AMANDA-Julia-2010}%
  \BibitemOpen
  \bibfield  {author} {\bibinfo {author} {\bibfnamefont {R.}~\bibnamefont
  {Abbasi}} \emph {et~al.} (\bibinfo {collaboration} {IceCube Collaboration}),\
  }\href {\doibase 10.1016/j.astropartphys.2010.05.001} {\bibfield  {journal}
  {\bibinfo  {journal} {Astropart. Phys.}\ }\textbf {\bibinfo {volume} {34}},\
  \bibinfo {pages} {48} (\bibinfo {year} {2010}{\natexlab{b}})}\BibitemShut
  {NoStop}%
\bibitem [{\citenamefont {{Abbasi}}\ \emph {et~al.}(2011)\citenamefont
  {{Abbasi}} \emph {et~al.}}]{Diffuse-Sean-2011}%
  \BibitemOpen
  \bibfield  {author} {\bibinfo {author} {\bibfnamefont {R.}~\bibnamefont
  {{Abbasi}}} \emph {et~al.} (\bibinfo {collaboration} {IceCube
  Collaboration}),\ }\href {\doibase 10.1103/PhysRevD.84.082001} {\bibfield
  {journal} {\bibinfo  {journal} {Phys. Rev. D}\ }\textbf {\bibinfo {volume}
  {84}},\ \bibinfo {pages} {082001} (\bibinfo {year} {2011})}\BibitemShut
  {NoStop}%
\end{thebibliography}
\end{document}